\documentclass{article}
\usepackage{amssymb}

\usepackage{graphicx}
\usepackage{amsmath}

\input{tcilatex}

\begin{document}

\title{Remarks on scaling properties inherent to the systems with hierarchically
organized supplying network}
\author{V.Gafiychuk{$^a$}, I.Lubashevsky{$^b$}, A.Stosyk{$^a$} \\
{$^a$}Institute of applied problems of mechanics and mathematics\\
National Academy of Science of Ukraine.\\
{$^b$}Institute of general physics \\
Academy of Science of Russian Federation.}
\date{}
\maketitle

\begin{abstract}
We study the emergence of a power law distribution in the systems which can
be characterized by a hierarchically organized supplying network. It is
shown that conservation laws on the branches of the network can, at some
approximation, impose power law properties on the systems. Some simple
examples taken from economics, biophysics etc. are considered.
\end{abstract}

The existence of power law distribution in different systems of nature is a
fascinating property still waiting for sufficient explanation. Despite a
very long history of investigation of different scaling laws in systems of
outworld this problem is very attractive up to the present time. A high
interest to this problem is associated primarily with the fact that the
systems, different in their nature, possess the similar power law
distributions. This in its turn indicates that complex systems which consist
of tremendous amount of objects have universal behavior, whose investigation
gives the possibility to penetrate deeply into the essence of organization
of nature.

V.Pareto (1897, \cite{p897}) seems to be first who discussed the power law
distributions in wealth of the individuals in a stable economy. Since that
time more than one century passed and the scaling properties are already
discovered in many systems. Now we meet power law distributions in
turbulence \cite{k62,ll}, in biological systems \cite{s99}, economics and
finance \cite{ms97,ott99,wmw99,nzipf}, social phenomena \cite{zhang},
linguistic phenomena \cite{z49}, systems which can be characterized by
network organization like river network \cite{dr99,ss97}, etc.

In this paper we single out a certain subclass of systems widely met in the
nature which obey power law. Main characteristic feature of such systems is
their hierarchical organization. The hierarchical organization allows such
system to be divided into sets of subsystems (which are called levels)
involving many elements that are similar in their properties. The elements
of the different levels are substantially different in their characteristics.

We consider systems that can exist only due to the permanent flow through
the system of some transport agents joining different subsystems and
elements into a whole organization. Such interconnection of systems elements
is formed by hierarchically organized supplying network. The specific
function of the network is to drain the systems or supply them with some
transport agents needed for systems activity. The characteristic feature of
such system consists of conservation laws which should be fulfilled on each
hierarchy level. This denotes that total flow of the transport agents
through the system is constant.


We represent the system under consideration by a simple model schematically
drawn on Fig.~(\ref{F.1},a). Each hierarchy level $i$ consists of set of $%
n_{i}$ independent elements. All these elements take part in transferring
transport agent flow from one level to another, namely taking it from the
elements belonging to the lower level, and transferring it to the elements
of higher level and 
\begin{equation*}
n_{i+1}>n_{i}>n_{i-1};\;\;\forall i.
\end{equation*}

Due to the transport agent flow being constant through each hierarchy level $%
i$ the total flow $X_{\text{\textrm{total}}}$ is constant and 
\begin{equation}
X_{\text{\textrm{total}}}=\sum_{j}X_{i,j}\thickapprox n_{i}X_{i},\;\;\forall
i,  \label{s3}
\end{equation}
here $n_{i}$-is the number of system elements on the $i$-th hierarchy level, 
$X_{i,j}$-the flow through the $j$-th element of the $i$-th hierarchy level, 
$X_{i}$-the average over level flow through each element of level $i$. It
was supposed that all the elements of the same hierarchy have similar
characteristics and it is possible to sum up their flows directly. In the
framework of this article we pay attention to the fact that the power law
dependence is inherent to different natural systems and their occurrence is
naturally connected to the function of systems and is the intrinsic property
of the systems.

The flow $X_{i}$ though each element is determined by characteristics of the
properties of the element and system organization. We denote all these
characteristics by some parameters $A_{i}$. Namely these parameters are
observable features of systems under consideration. It is conventional way
to investigate power law distribution of these parameters. In this case from
(\ref{s3}) we get an obvious relation 
\begin{equation}
n_{i}\backsim 1/X_{i}(A_{i}).  \label{s4}
\end{equation}
We will touch some examples of model to determine $A_{i}$.

Now consider a simple example taken from economics. As the economic systems
exist only with the streams of commodity and money, they possess all the
above-listed properties.

The number of parameters, which define the economic market state, at first
glance seems to be actually endless because there is a tremendous amount of
goods on the market. At the same time a relatively small number of raw
materials in the industry shows that there should be a large hierarchical
systems in the market structure. So it is possible to present an economic
system as a hierarchical system which hierarchy levels consist of the firms
of different power. On the first level there are firms processing row
materials, on the last level -- the shops of retail trade. In this context
we point out that after reaching the consumers goods the flows transform
into money flows are going in the opposite direction. The schematic
representation of such system is shown in fig.~(\ref{F.1},b). We shall
consider the situations, when the market of the considered goods exist, i.e.
all the streams should be more than zero.

For simplicity we confine our consideration to some commodity market, in
which producers, dealers and consumers of some kinds of goods will form a
common connected network. For example, the market of steel or meat items and
the market of furniture can interact with each other mainly through the
changes in financial states of the whole set of the consumers. In this case
we can consider some branch of industry as a single one which practically
does not interact with any other branches of industry~\cite
{gl99,gl98,gl97,gl95}. \footnote{%
It should be noted that this restriction is not so strong. Main conclusion
of this article will be true for the whole economic system (for example, the
economic system of the state), including many sets of material flows twisted
with each other. Indeed, in this case we can expend total material flow onto
the basis of different material flows. For each of them situation will be
practically the same as listed below.}

The level $i$ of this system is a set of $n_{i}$ independent firms, bringing
out the products of some sort, which buy the products from firms of the
lower level of the hierarchy, and sell the results of their activity to the
firms of the higher level. Firms, forming the common level $i$, are
considered as identical from the viewpoint of their power. The bottom of the
system (the firms of the first level) is formed by the firms, obtaining and
processing raw material. The market participants of the last level are the
``points'' of retail trade, supplying the consumers with required goods.

The level of the firm production on each level $i$ is measured in the units
of the initial raw material flow $X_{i}$, ``flowing'' through the given
firm. It should be noted, that in this system of units the value $X$ having
dimensionality (\textit{material})/(\textit{time}). Particular
interconnection between different firms can occur and disappear during the
formation and evolution of the market under consideration. Their interaction
is governed by trade with firms belonging to the neighbor levels.

\FRAME{ftbpFU}{3.2067in}{2.6256in}{0pt}{\Qcb{Representation of the flows
between levels in the systems with hierarchically organized supplying
structure. In Fig. a: segments correspond to firms and arrows to flows; the
volume of the segments and arrows represent hierarchical organization of the
system; Fig. b corresponds to the schematic representation of market
structure in the tree form.}}{\Qlb{F.1}}{F.1}{\special{language "Scientific
Word";type "GRAPHIC";maintain-aspect-ratio TRUE;display "USEDEF";valid_file
"F";width 3.2067in;height 2.6256in;depth 0pt;original-width
5.5037in;original-height 4.4987in;cropleft "0";croptop "1";cropright
"1";cropbottom "0";filename '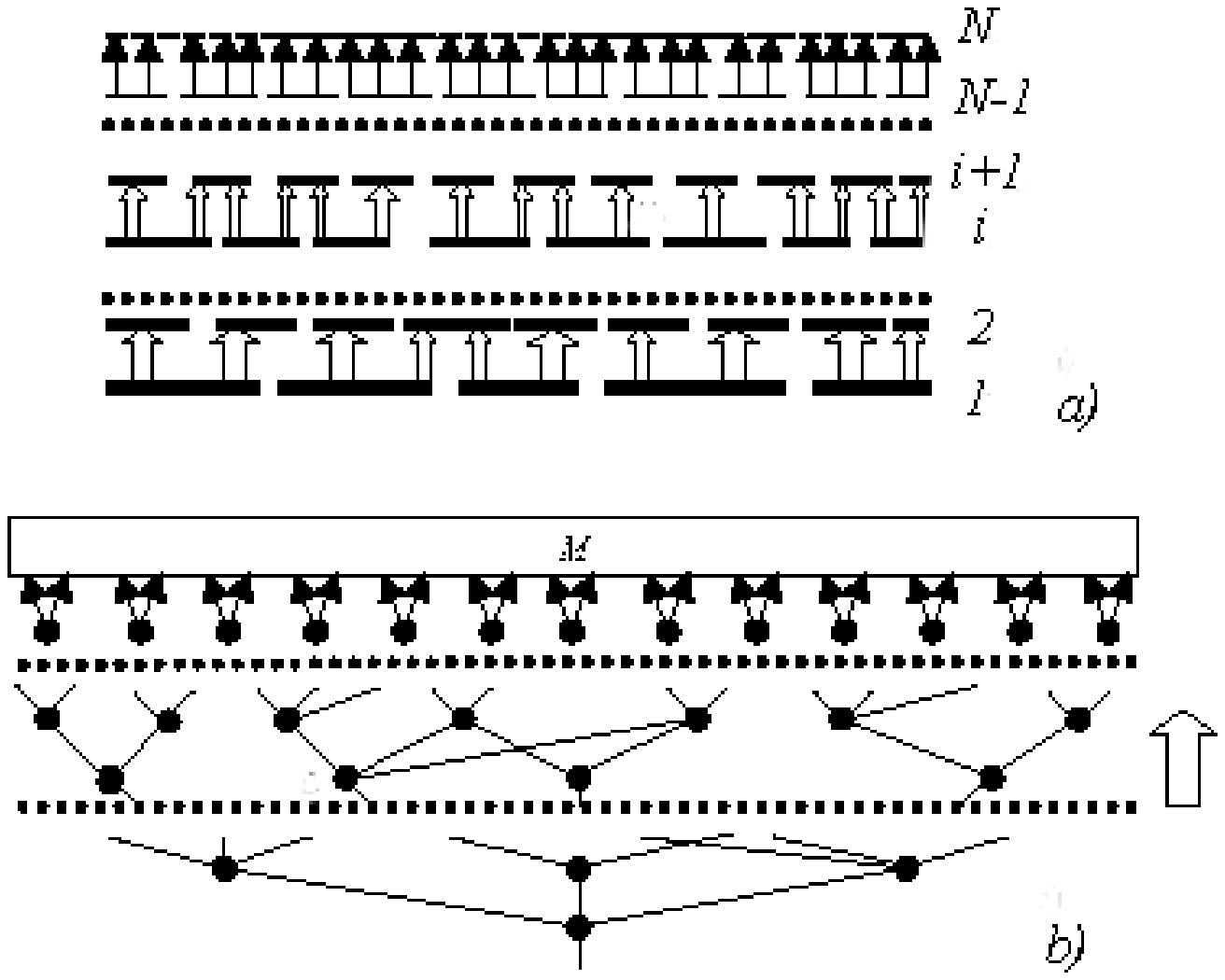';file-properties "XNPEU";}}

For simplicity, consider some hierarchical system which can be represented
as a homogeneous tree. Each firm is represented by the node on the tree. One
branch enter this node and $a$ branches go out. Such representation denotes,
that the given firm connected with trade relations with one firm of the
lower level and \ $a\;$firms of the higher level. It is assumed that tree
consist of $N$ levels. In this case number of firms on $i$-th level - $n_{i}$
- equal 
\begin{equation}
n_{i}=a^{i-1}.  \label{frag1}
\end{equation}
The fraction of the system, which make up the $i$-th level is

\begin{equation}
p(i)=\frac{n_{i}}{S_{N}},  \label{frag2}
\end{equation}
where $S_{N}$ - number of nodes in the tree (firms in our system)

\begin{equation}
S_{N}=\sum_{i=1}^{N}n_{i}=\frac{a^{N}-1}{a-1}.  \label{frag6}
\end{equation}

The fraction $p(i)$ may be regarded as the probability that chosen at random
firm belong to the $i$-th level. Cumulative probability defined as the
probability that chosen at random firm belong to one of levels from the 1-st
to the $i$-th 
\begin{equation}
P(i)=\frac{S_{i}}{S_{N}}=\frac{an_{i}-1}{a^{N}-1},  \label{frag7}
\end{equation}
where $S_{i}=\sum_{k=1}^{i}n{_{k}}$.

As it was mentioned above $n_{i}=X_{\text{\textrm{total}}}/X_{i}$ due to
this fact probability to choose node with flow $x$ less or equal to $X_{i}$
(cumulative distribution function) is 
\begin{equation}
P(x\geqslant X_{i})=\frac{a(X_{\text{\textrm{total}}}/X_{i})-1}{a^{N}-1}%
\backsim \frac{1}{X_{i}}.  \label{frag11}
\end{equation}

Finally, it is important to express flow $X_{i}\;$using the dependence of
the parameter $A_{i}$ from the flow $X_{i}$.$\;$There is a question what the
parameter $A_{i}$\ means. For example let us consider the parameter $A_{i}\;$%
as a firm's profit $\pi _{i}$.\ It is natural to suppose that profit of each
firm is proportional to the material or financial flow through the firm,
namely $\pi _{i}\backsim X_{i}^{\alpha }$ $(\alpha >0)$.$\;$ Then we can
write cumulative probability distribution of firm's profit for $n_{i}\gg 1$

\begin{equation}
P(\geqslant \pi _{i})\backsim \frac{1}{\pi _{i}^{1/\alpha }}.  \label{frag16}
\end{equation}

In the most intuitive case $\alpha =1$ and we obtain Zipf's law distribution
as it was observed in \cite{ott99}. It seems to be reasonable to note that
distribution of the firms from their income analyzed in the \cite{ott99} is
a direct consequence of the hierarchical organization of the firms into
corresponding industrial branches, which are connected to each other by
material and financial flows.

As an another typical example of the systems under consideration we may
regard living tissue where blood, flowing through the vascular network,
which involves arterial and venous beds, supplies cellular tissue with
oxygen, nutritious products, etc. At the same time blood withdraws carbon
dioxide and products resulting from life activities of the cellular tissue.
Both the arterial and venous beds being of the tree form contain a large
number of hierarchy levels and are similar in structure. A similar situation
takes place in respiratory systems where oxygen going through hierarchical
system of bronchial tubes reaches small vessels (capillaries).

A microcirculatory bed of living tissue can be reasonably regarded as a
space-filling fractal, being a natural structure for ensuring that all cells
are serviced by capillaries \cite{WBE97,we}. The vessel network must branch
so that every small group of cells, referred below to as ``elementary tissue
domain'', is supplied by at least one capillary. In this case a vessel
network generated by arteries should contain a sufficiently large number of
hierarchy levels At each level $i$ of the vascular network the tissue domain
supplied by a given microcirculatory bed as a whole can be approximated by
the union of the tissue subdomains whose mean size is about the typical
length $l_{i}$ of the $i$-th level vessels. Thus, the individual volume of
these subdomains is estimated as $V_{i}\sim l_{i}^{3}$. From this simple
consideration immediately follows the sufficiently obvious power law
distribution for vascular network. Indeed, for the space-filling vascular
network this artery supplies with blood the tissue region of volume about $%
l_{i}^{3}$ and, so, under normal conditions the blood flow rate $X_{i}$ in
it should be equal to 
\begin{equation}
X_{i}\approx jl_{i}^{3},  \label{s6}
\end{equation}
where $j$ is the blood perfusion rate (the volume of blood, flowing through
the tissue domain of volume unit\ per time unit) assumed to be the same at
all the points of the given microcirculatory bed. In living tissue the ratio 
$l_{i}/a_{i}$ takes a certain fixed value, $l_{i}\approx \mathrm{constant}%
\cdot a_{i}$ and, thus, $X_{i}\approx \mathrm{constant}^{\prime }\cdot
a_{i}^{3}$. Due to the blood conservation at branching nodes the scaling law
of such system of vessels controlling finally the blood flow redistribution
over the microcirculatory beds can be represented as 
\begin{equation}
n_{i}\backsim const/a_{i}^{3}\backsim const/l_{i}^{3}  \label{s7}
\end{equation}

In the same way we can obtain approximate power law dependance for river
network organization. Recapitulating practically one-to-one the explanation
listed above, for the hierarchically river network as two-dimensional
space-filling fractal we can write 
\begin{equation*}
X_{i}\approx jl_{i}^{2}.
\end{equation*}
In this case from equation (\ref{s3}) we get 
\begin{equation}
n_{i}\backsim const/l_{i}^{2}\backsim const/l_{i}^{2}.  \label{s8}
\end{equation}
The obtained power law dependence is approximately valid for qualitative
understanding of scaling properties so as the results obtained in framework
of this approach correspond to that one, obtained by more rigorous
consideration \cite{dr99}.

From the described examples we can develop a general mathematical model of
the system under consideration consisting from the distributed basic medium $%
\mathbb{M}$ and the transport hierarchical network ${S}$ (Fig.~\ref{F.1},b).
The transport network can supply basic medium by some transport agent or can
drain basic medium. The medium $\mathbb{M}$ is a $d$ - dimensional
homogeneous continuum. The transport network can be reasonably regarded as a
space-filling fractal embedded in $d$-dimensional homogeneous continuum. 
\begin{equation*}
X_{i}\approx jl_{i}^{d}
\end{equation*}
where $j$ is the flow of transport agents (the volume of transport agent
through some domain of unit volume per unit time) assumed to be the same at
all the points of the given bed. In this case $n_{i}\approx const/l_{i}^{d},$
and the dimensionality of homogeneous continuum should be determined. In the
given examples such basic medium has dimensions $d=1$, $d=3$ and $d=2$
respectively to (\ref{s6}),(\ref{s7}),(\ref{s8}). In general case such
dimensions can be noninteger and their properties seem to determine the
properties of basic medium $\mathbb{M}$ as a whole.

\end{document}